\documentclass[aps,prl,twocolumn,showpacs,groupedaddress,floatfix]{revtex4}
\usepackage{graphicx,graphics}
\begin{document}

%%%%% shortcuts %%%%%%
\def\deg{^\circ}
\newcommand{\ind}[1]{_{\mbox{\scriptsize#1}}}     % indices
\newcommand{\indt}[1]{_{\mbox{\tiny#1}}}          % tiny indices
\newcommand{\supind}[1]{^{\mbox{\scriptsize#1}}}  % superscript
\newcommand{\unit}[1]{\, \mbox{#1}}               % format for units
\newcommand{\imag}{\mbox{i}}                      % sqrt(-1) non-italic
\newcommand{\imind}{\mbox{\scriptsize{i}}}        % sqrt(-1) non-italic for indices
%%%%% shortcuts %%%%%%

\title{
Surface Roughness and Hydrodynamic Boundary Conditions}

\author{Olga~I.~Vinogradova}
\email[Corresponding author: ] {vinograd@mpip-mainz.mpg.de}
\affiliation{Max Planck Institute for Polymer Research, Ackermannweg
10, 55128 Mainz, Germany} \affiliation{A.N.Frumkin Institute of
Physical Chemistry and Electrochemistry, Russian Academy of
Sciences, 31 Leninsky Prospect, 119991 Moscow, Russia}

\author{Gleb~E.~Yakubov\footnote{Present address:
Corporate Physical and Engineering Sciences, Unilever, R\&D Colworth
Sharnbrook, Bedfordshire MK44 1LQ, England, UK}} \affiliation{Max
Planck Institute for Polymer Research, Ackermannweg 10, 55128 Mainz,
Germany} \affiliation{A.N.Frumkin Institute of Physical Chemistry
and Electrochemistry, Russian Academy of Sciences, 31 Leninsky
Prospect, 119991 Moscow, Russia}

\date{\today}

\begin{abstract}
We report results of investigations of a high-speed drainage of thin
aqueous films squeezed between randomly nanorough surfaces. A
significant decrease in hydrodynamic resistance force as compared
with predicted by Taylor's equation is observed. However, this
reduction in force does not represents the slippage. The measured
force is exactly the same as that between equivalent smooth surfaces
obeying no-slip boundary conditions, but located at the intermediate
position between peaks and valleys of asperities. The shift in
hydrodynamic thickness is shown to be independent on the separation
and/or shear rate. Our results disagree with previous literature
data reporting very large and shear-dependent boundary slip for
similar systems.
\end{abstract}
\pacs {82.70.Dd, 83.80.Qr, 82.70.-y}

%\keywords{}

\maketitle

%\section{Introduction}

It has been recently well-recognized that the classical no-slip
boundary condition~\cite{lamb.h:1932}, which has been applied for
more than a hundred years to model macroscopic experiments, is often
not applicable at the submicro- and especially nanoscale. Although
the no-slip assumption seems to be valid for molecularly smooth
hydrophilic surfaces down to a
contact~\cite{chan.d:1985,israelachvili.j:1986,klein.j:1993,vinogradova:03,charlaix.e:2005},
it is now clear that this is not so for the majority of other
systems. The changes in hydrodynamic behavior are caused by an
impact of interfacial phenomena, first of all hydrophobicity and
roughness, on the flow. A corollary from this is that a theoretical
description based on the no-slip condition has to be corrected even
for simple liquids. What, however, still remains a subject of hot
debates is that how to correct the flow near the interface, and what
would be the amplitude of these corrections.

The simplest and the most popular way to model the flow is to use a
slip-flow approximation~\cite{deGennes.pg:1985}, which assumes that
the slip velocity at the solid wall is proportional to the shear
stress, and the proportionality constant is the so-called slip
length. Such an assumption was justified theoretically for smooth
hydrophobic
surfaces~\cite{vinogradova:99,barrat:99,andrienko.d:2003} and was
confirmed by recent surface force apparatus
(SFA)~\cite{charlaix:01,charlaix.e:2005} and atomic force microscope
(AFM)~\cite{vinogradova:03} dynamic force experiments. Despite some
remaining controversies in the data and amount of slip
(cf.~\cite{granick.s:2003}), a concept of a hydrophobic slippage is
now widely accepted.

For rough surfaces a situation is much less clear both on the
theoretical and experimental sides. One point of view is that
roughness decreases the degree of
slippage~\cite{granick.s:2003,granick:02,pit:2000}, unless the
surface is highly hydrophobic, so that trapped nanobubbles are
formed to accelerate the
flow~\cite{vinogradova.o:1995,cottin_bizonne.c:2003.a}. An opposite
conclusion is that roughness generates extremely large
slip~\cite{bonaccurso.e:2003}.

We believe our letter entirely clarifies the situation with flow
past rough surfaces, highlights  reasons for existing controversies,
and resolves apparent paradoxes.

%\section{Experiment}

\begin{figure}
\includegraphics[width=6cm,clip]{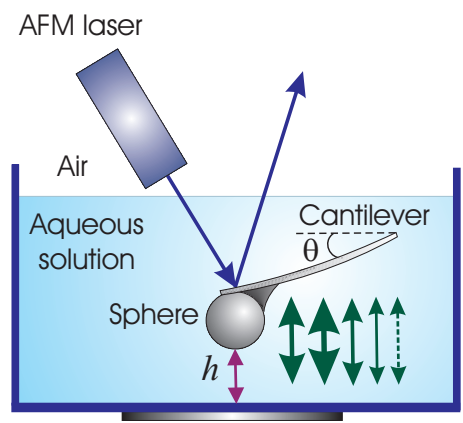}
\caption{Schematic of the dynamic AFM force experiment.}
\label{fig1}
\end{figure}

\begin{figure}
\includegraphics[width=6cm,clip]{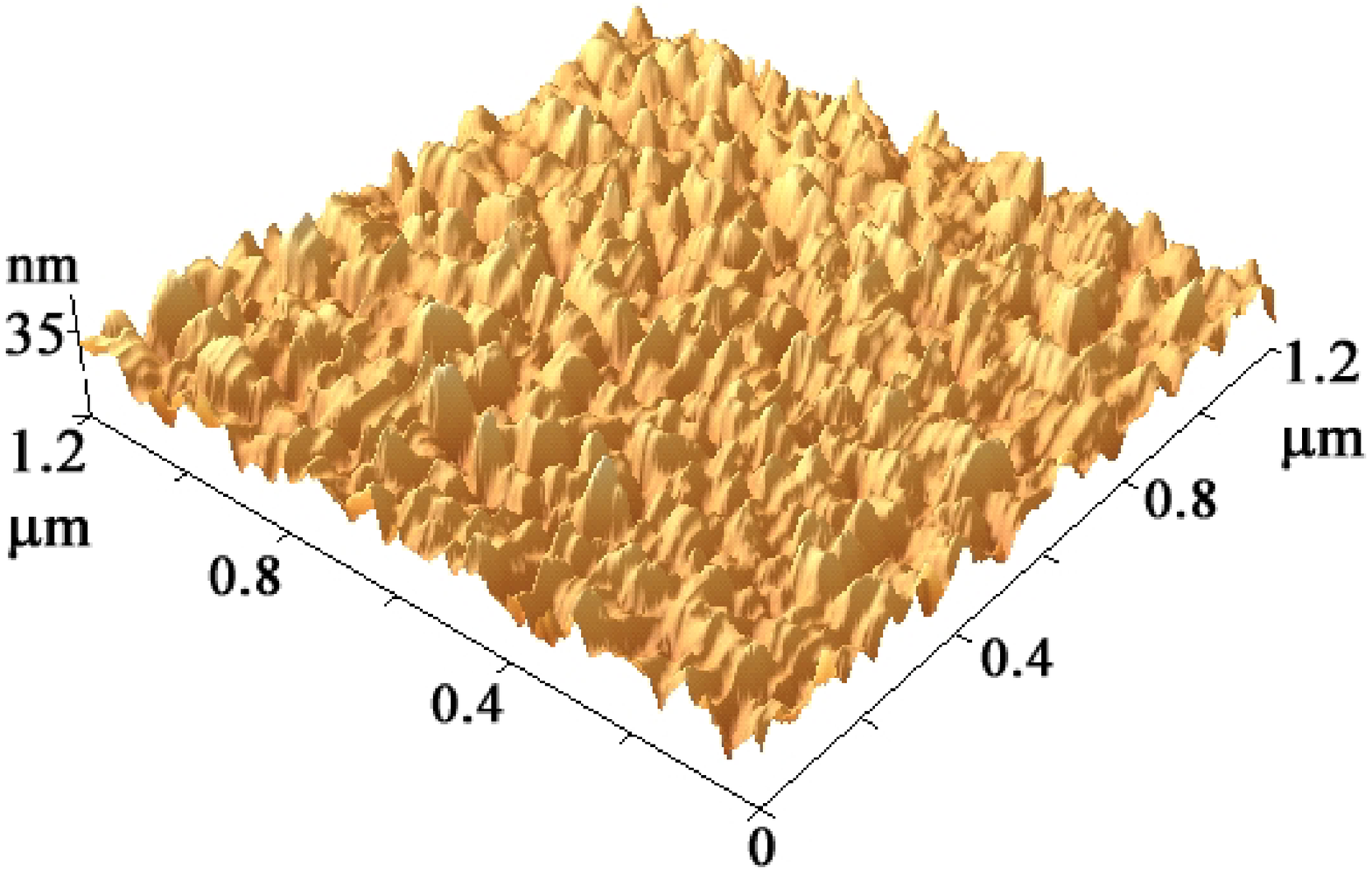}\\
\includegraphics[width=6cm,clip]{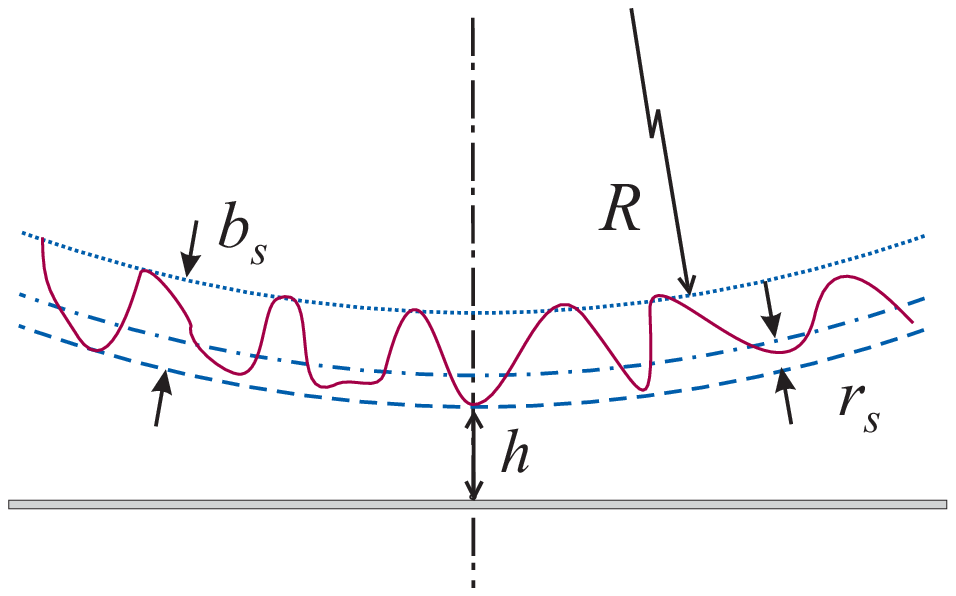}
\caption{An AFM image of an apex of a gold-coated sphere (top) and a
schematic representation of rough sphere-plane interaction near the
point of a contact (bottom).} \label{fig2}
\end{figure}

We use a specially designed home-made AFM-related
setup~\cite{ecke.s:2001,vinogradova:01,yakubov:00} to perform
dynamic force experiments on a nanoscale (Fig.~\ref{fig1}). Glass
spheres of radius $R \sim 20 \pm 2$ $\mu$m (Duke Scientific,
PaloAlto, CA) were attached with UV glue on the top of the
rectangular tipless cantilever (length $L = 450$ $\mu$m, width $w =
52$ $\mu$m, spring constants $k \sim 0.22$ N/m). The spheres were
then coated with a gold layer (50 nm) using a layer of chromiun (3
nm) to promote adhesion. For the planar substrate we used a silicon
wafer coated with a gold layer (50 nm). Both surfaces were then
treated with a 1 mM solution of 11-amino-undecane thiol
(SH-(CH$_2$)$_{11}$-NH$_2$) for 24 hrs to produce a chemically bound
SAM. The advancing and receding water contact  angle on the
thiolated planar surfaces were measured with a commercial setup
(Data Physics, Germany), and were found to be ($69 \pm 1$)$^{\circ}$
and ($63 \pm 1$)$^{\circ}$, respectively. These values are close to
those for surfaces used in~\cite{craig:01}. Imaging of thiolated
surfaces with a regular AFM tip revealed the root-mean-square
roughness over a $1 \mu{\rm m} \times 1 \mu{\rm m}$ is in the range
10-11 nm for a sphere and 0.5-0.8 nm for a plane. The maximum
peak-to-valley difference is less than 45 nm for a sphere
(Fig.~\ref{fig2}) and less than 2.5 nm for a plane. This (smooth
against rough) geometry of configuration allows one to avoid large
scatter in data at separations of the order of or smaller than the
roughness size. Such a scatter would be inavoidable for two rough
surfaces (depending whether the rough sphere is falling on a tip or
in a valley of a rough plane).
 Cantilevers were then fixed in a holder with the
variable tilt angle, and the intermediate position with the angle
$\theta = (10 \pm 2)^{\circ}$  was chosen. A planar substrate was
placed onto the bottom of a Teflon cuvette, which was filled with
1mM NaCl (99.99\%, Aldrich) aqueous solutions. Water for solutions
was prepared using a commercial Milli-Q system containing
ion-exchange and charcoal stages. The deionized water had a
conductivity less than $0.1 \times 10^{-6}$ S/m and was filtered at
0.22 $\mu$m. To measure force-versus-position curves the cuvette was
moved towards the particle with a 12 $\mu$m range piezoelectric
translator (Physik Instrumente, Germany). This translator was
equipped with integrated capacitance position sensors, which allows
to avoid any creep and hysteresis and provided the position with an
accuracy of 1 nm in closed-loop operation. During the movement the
deflection of the cantilever was measured with an optical lever
technique. Therefore the light of a laser diode (1.5 mW, 670 nm) was
focused onto the back of the gold coated cantilever. After
reflection by a mirror, the position of the reflected laser spot was
measured with a position sensitive device (United Detectors, UK,
active area $30 \times 30$ mm$^2$). The total force was calculated
by multiplying instantaneous cantilever deflection with the spring
constant. The distance $h$ between surfaces was calculated by adding
the piezo displacement to the deflection of the cantilever, so that
$h=0$ corresponds to the contact [of tips of sphere's asperities
with a plane] (see Fig.~\ref{fig1}). We stress, that since our plane
is smooth, we have no ambiguity in determining this zero of
separation.

%\section{Theoretical calculations of film thinning}

The AFM force balance incorporates both (concentrated) force on the
sphere and the drag on the cantilever~\cite{vinogradova:03}
(Fig.~\ref{fig1}), so that the total force measured in the AFM
dynamic experiment is
\begin{equation}
\label{eq:total_force} F_t = \frac{-1 + 3 \cos {\theta}}{2} \left(F_s + F_h \right) + F_c,
\end{equation}
where  $F_s$ and $F_h$ are the surface and the hydrodynamic forces, correspondingly, acting on a sphere, and $F_c$  is the force due to distributed hydrodynamic drag on the cantilever.

The total force $F_t$  is proportional to the instantaneous
deflection of the end of the spring from its equilibrium position
multiplied by the spring constant $k$~\cite{chan.d:1985}
\begin{equation}
\label{eq:force_t} F_t = k \left(h - (h_0 + v t) \right),
\end{equation}
where $h_0$  is the initial separation between surfaces, and  $v$ is
the driving speed of piezo (negative speed corresponds to approach).

The hydrodynamic force $F_h$  between a sphere and a plane can be
written as~\cite{vinogradova:95}
\begin{equation}
\label{eq:force_h} F_h = - \frac{6 \pi \mu R^2}{h} \frac{d h}{d t} f^{\ast},
\end{equation}
where $\mu$ is the dynamic viscosity, $-d h / d t$ is the relative
velocity of the surfaces. To finalize the description of $F_h$ we
should define the expressions for a correction function $f^{\ast}$,
which will be discussed later.
  Note that when
$f^{\ast} = 1$, Eq.~\ref{eq:force_h} transforms to famous Taylor's
formula (which, however, never appeared in any of G.I.Taylor's
publications as discussed in~\cite{horn:00}).

The drag force on a cantilever $F_c$  is given
by~\cite{vinogradova:03}
\begin{equation}
\label{eq:force_c} F_c = - \frac{v \mu L}{2}\left[ \left(\frac{w}{2 R + h}\right)^3 \gamma^{\ast} + B \right]
\end{equation}
with
\begin{equation}
\label{eq:gamma_ast} \gamma^{\ast} = \gamma \left[ 1 - \frac{3 \gamma}{2} + 3 \gamma^2 - 3 \gamma^3 \ln \left(1 + \frac{1}{\gamma} \right) \right],
\end{equation}
where  $\gamma = (2 R + h) / (L \sin {\theta})$. Here $B$ is a
constant, which reflects the contribution from the  Stokes flow to
the cantilever deflection, and represents an adjustable
(dimensionless) parameter
 to the model.

%\section{Results and discussion}

The surface force $F_s$ is assumed to  be unaffected by the relative
motion of surfaces, and was taken to be the equilibrium force being
a function of only $h$. It was obtained from low speed (below 1
$\mu$m/s) force measurements. At distances larger than 20-25 nm no
interaction was detected. In other words, no electrostatic
contribution can be seen despite relatively large (9.6 nm) Debye
length in 1 mM NaCl solution. This suggests that the surfaces are
uncharged.  Similar observations have been made before for some
other classes of thiols~\cite{ederth.t:1998}. The range of the jump
distance was always $15 \pm 2$ nm. The contribution of contact
deformation to the jump distance was estimated using the
experimental values for the pull-off force $0.70 \pm 0.05$ mN/m
(with the correction to the hydrodynamic interaction) and the values
of Young's modulus of the UV glue (3 GPa), as it was the softest
material in our system. We also ignore the possible plastic
flattening of the gold (50 GPa) asperities. For these parameters the
central displacement due to a contact deformation is of the order of
0.1 nm, so that it can safely be ignored. We have fitted the
experimental results by assuming $F_s = -A R/6 h^2$, taking the
Hamaker constant, $A$, as an adjusting parameter. The value $A = 5.4
\times 10^{-20}$ J is obtained from fitting and was further used in
all calculations.

\begin{figure}
\includegraphics[width=7.5cm,clip]{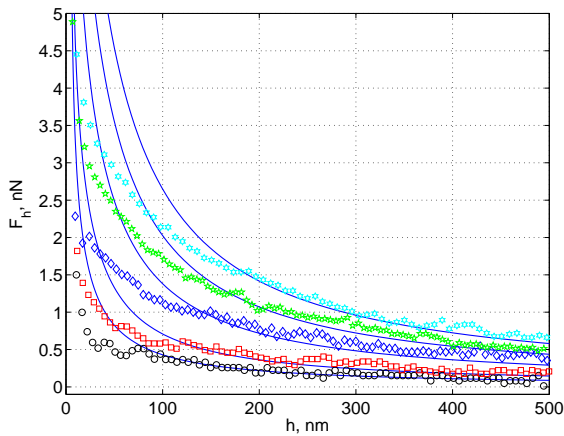} %
\caption{Hydrodynamic force acting on a sphere at high speed of
approach. Only each 2nd  point is shown. From bottom to top are data
(symbols) obtained at the driving speed -6 $\mu$m/s, -10 $\mu$m/s,
-20 $\mu$m/s,   -30 $\mu$m/s,   and -40 $\mu$m/s. Solid curves show
the calculation results obtained for the same speed, by assuming
$f^{\ast} = 1$.} \label{fig3}
\end{figure}

Fig.~\ref{fig3} shows the hydrodynamics resistance force calculated
by subtracting $F_s$ and $F_c$ from the total force measured.
Theoretical curves obtained by a numerical solution of differential
Eq.(\ref{eq:total_force}) in the assumption $f^{\ast} = 1$ are also
included. Note that cantilever contribution was found to be neither
small or negligible. The adjusting parameter $B$ reflecting
Stokes-like flow on a cantilever was about 30-34 for our geometry of
configuration and slightly varied from one experiment to another.
Measured $F_h$ is much smaller than predicted by Taylor's theory.
The deviations from theory are clearly seen at distances 100-200 nm
depending on the driving speed. One can conclude therefore that
there is clear impact of roughness on the film drainage.

\begin{figure}
\includegraphics[width=7.5cm,clip]{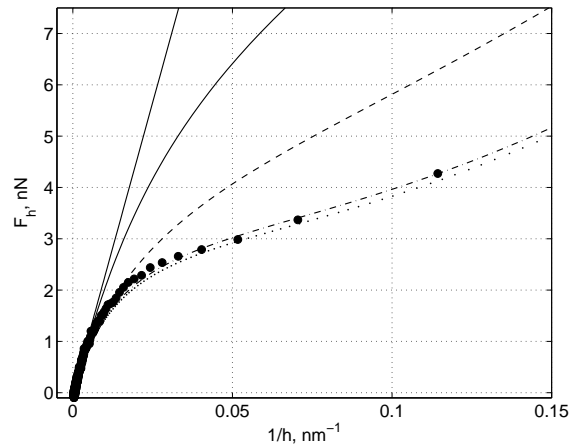} %
\caption{Hydrodynamic force versus inverse of separation obtained at
driving speed -30 $\mu$m/s. Symbols represent experimental data.
Solid curves show the theoretical results obtained within Taylor's
formula. Upper line shows the calculations results in the assumption
of a constant speed of approach. Lower curve corresponds to the real
approaching speed. Dashed curve is the calculation results within
the model described by Eq.~\ref{slip} with $b_s = 48$ nm, and dotted
curve - by Eq.~\ref{shift} with $r_s = b_s = 48$ nm. The
approximation Eq.~\ref{shift} with $r_s = 45$ nm provides the best
fit to the data. } \label{fig4}
\end{figure}

In Fig.~\ref{fig4} the hydrodynamic force measured at driving speed
$-30 \mu$m/s is compared with Taylor's equation (solid curves). The
force is plotted versus the inverse of the surface separation. Note
that even Taylor's equation does not result in a linear plot. This
would be expected only at a constant approach velocity, which is not
the case due to forces acting on the sphere. However these, caused
by decrease of $dh/dt$, deviations from linearity are much smaller
than required to fit experimental data. The general analytical
solution for rough interfaces does not exist in the literature
probably since such a problem is outside the scope of a lubrication
approximation. Below we analyze some approximate models.

We have calculated the force expected between smooth slippery
surfaces. The slip length is assumed to be roughly equal to zero for
a sphere and to the height of asperities $b_s$ (as defined in
Fig.~\ref{fig2}) for a sphere, which is close to its definition. In
this case the correction for slippage takes the
form~\cite{vinogradova:95}
\begin{equation}
f^{\ast} = \frac{1}{4} \left( 1 + \frac{3 h}{2 b_s}\left[ \left( 1 +
\frac{h}{4 b_s} \right) \ln \left( 1 + \frac{4 b_s}{h} \right) - 1
\right]\right).\label{slip}
\end{equation}
An improved fit to the data is obtained when slip is permitted
(dashed curve in Fig.~\ref{fig4}). For this speed of approach the
best fit to Eq.~\ref{slip} was found with $b_s = 48$ nm.
Eq.~\ref{slip} provides an excellent fit at distances larger than
$\sim$ 100 nm. However, since data still deviate from theoretical
predictions at smaller separations, it is clear that slippage does
not necessarily represent the roughness.

Therefore, we further assumed that stick boundary conditions remain
valid, but are applied to a surface defined at a distance $r_s$
towards a sphere:
\begin{equation}
f^{\ast} = \frac{h}{h + r_s}\label{shift}
\end{equation}
By adjusting the value of $r_s$ we found that the hypothesis of a
shift in separation gives a perfect coincidence between data and
theoretical predictions (dash-dotted curve in Fig.~\ref{fig4},
obtained with $r_s=45$ nm). We remark and stress that
Eq.~\ref{shift} provides an excellent description of the data even
when $h$ is much smaller than $r_s$. [Another point to note is that
the values of $r_s$ seem to be slightly larger than would be
expected from the AFM imaging, which is likely connected with the
fact that the AFM tip has a finite size and cannot go into grooves].

We thus conclude that the description of flow near rough surfaces
has to be corrected, but this is not the relaxation of no-slip
boundary conditions, but the correction for separation with its
shift to larger distances within the asperities size. It has to be
stressed that similar ideas were justified theoretically at
macroscale for a shear flow along periodically corrugated
wall~\cite{richardson:71} and recently for a far-field motion of a
sphere towards such a wall~\cite{lecoq.n:2004}. They are also
consistent with molecular-dynamic simulations on simple model
systems~\cite{barrat:94}. Here for the first time we provided
accurate experimental data showing that the concept can be applied
for a randomly rough wall and at the nanoscale, down to a contact.
Note that Eq.~\ref{shift} has not received so far any theoretical
justification for short, i.e. of the order of or smaller than the
size of asperities, separations. This is probably because this
situation escapes from the framework of the lubrication
approximation since two length scales of the problem become
comparable. The applicability of Eq.~\ref{shift} at the short
distances probably reflects the fact that the height of the
roughness elements is much smaller than the sphere radius, so that
even when separation is getting small, the plane ``sees'' many
roughness elements at the same time and fluctuations are averaged
out. This hypothesis remains the subject for further theoretical
work.

\begin{figure}
\includegraphics[width=7.5cm,clip]{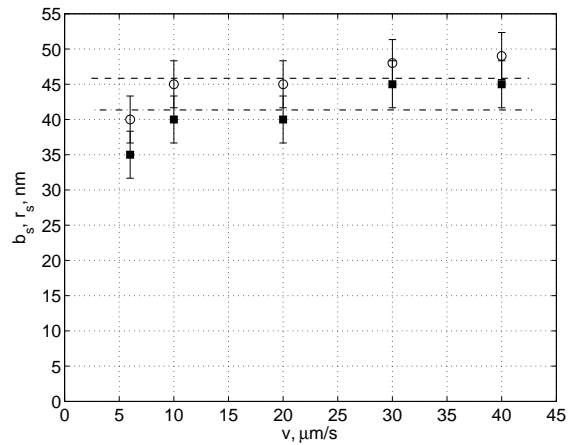} %
\caption{A hydrodynamic height of asperities $b_s$ (open circles)
and a hydrodynamic shift $r_s$ (filled squares) obtained at
different speed of approach. } \label{fig5}
\end{figure}

This conclusion remains valid for all driving speeds, and results
are summarized in Fig.~\ref{fig5}. One can see that $r_s$ remains
roughly the same at all speeds, although it shows some weak tendency
to increase with the rate of approach. However, this increase is
within the error of experimental data, so it remains an open
question for future research. The same remarks are true for $b_s$,
which is included for completeness since slip-model is applicable at
large separations.

Note that our results clarify the reason for apparently
contradictory reports suggesting that roughness
increase~\cite{pit:2000} and decrease~\cite{bonaccurso.e:2003} the
drag force. In our opinion these different conclusions only reflect
that in these papers the wall location was defined differently. It
follows from our results that if equivalent smooth surface is
defined as coinciding with the location of valleys of asperities
(which corresponds to experimental techniques used
in~\cite{pit:2000}), the measured force is larger than expected for
an equivalent surface (cf. dotted curve in Fig.~\ref{fig4}). If,
however, it is defined on the peaks of asperities (experimental
techniques used in~\cite{bonaccurso.e:2003}), then the measured
force is smaller (cf. dashed curve in Fig.~\ref{fig4}). Clearly,
both results~\cite{pit:2000,bonaccurso.e:2003}  physically mean that
roughness increase the dissipation in the system, and that an
equivalent surface is located at the intermediate position. Note,
that in~\cite{granick.s:2003}, which reports the increase in force
with roughness, the equivalent surface is also defined on the peaks
of asperities. We do not have any explanation of this result.

Our data and conclusions do not confirm results obtained with
similar, but ``smooth'' surfaces, where shear-rate dependent
slippage was detected~\cite{craig:01}. In our opinion, the reduction
in hydrodynamic resistance force might indicate that their
gold-coated thiolated sphere was in reality rougher than expected.
Reasons for a shear-dependency could probably be connected with some
errors in experimental determination of $dh/dt$ since the
piezotranslator used in~\cite{craig:01}, and later
in~\cite{bonaccurso.e:2003}, is not suitable for highly dynamic
force measurements due to its non-linearity. Another reason for a
significant rate-dependence is the use of binary
mixtures~\cite{craig:01}. Clearly, both confinement and shear might
lead to their stratification and a formation of a thin low viscosity
lubricant layer~\cite{andrienko.d:2003}. It is well known and has
been proven experimentally~\cite{horn:00} that such a layer leads to
rate-dependent phenomena (normally expected only at a very large
shear rate~\cite{thompson.pa:1997}) even at low speed. We suspect
that this effect of stratification of binary mixture, enhanced by
roughness, is responsible for a very large reduction in force
observed in~\cite{bonaccurso.e:2003}. Finally, we would like to
stress that since the force balance represents a differential
equation, even small $F_c$ could implicate the results by decreasing
$dh/dt$, and therefore $F_h$. The similar remark concern $F_s$,
which cannot be excluded from analysis. Both $F_s$ and $F_c$ are not
present in the force balance specified
in~\cite{craig:01,bonaccurso.e:2003}, which might cause further
inaccuracy in results.

In summary, by performing high-speed drainage experiment with
nanorough hydrophilic surfaces we demonstrate that their interaction
is similar to that between equivalent smooth surfaces located at
some position between peaks and valleys of asperities. Thus, our
results are in favor of no-slip assumptions for a hydrophilic
surface valid down to a contact.

%\section*{Acknowledgement}

This work was supported by a DFG through its priority program
``Micro and Nanofluidics'' (Vi 243/1-2). We thank E.Bonaccurso,
H.J.Butt, and V.S.J.Craig for discussion of details of their
experiments, and F.Feuillebois for helpful remarks on the
manuscript.

\bibliography{thiolb}
\end{document}